\documentclass[showpacs,preprintnumbers,amsmath,amssymb,aps,twocolumn,nofootinbib]{revtex4-1}

\usepackage{graphicx}
\usepackage{dcolumn}
\usepackage{color}

\usepackage{hyperref}
\hypersetup{%
colorlinks=true,%
linkcolor=blue,
citecolor=blue,
urlcolor=magenta
}
\newcommand{\sumint}{\mbox{$\sum$}\kern-2.7ex\int}

\def\gtsim{\mathrel{\hbox{\raise0.2ex
\hbox{$>$}\kern-0.75em\raise-0.9ex\hbox{$\sim$}}}}
\def\ltsim{\mathrel{\hbox{\raise0.2ex
\hbox{$<$}\kern-0.75em\raise-0.9ex\hbox{$\sim$}}}}

\newcommand{\dsla}{\partial\kern-1.3ex /~}

\begin{document}

\preprint{OCHA-PP-377}

\title{Electron electric dipole moment and electroweak baryogenesis in a complex singlet extension of the Standard Model with degenerate scalars}

\author{Chikako Idegawa$^{1}$}
\email{c.idegawa@hep.phys.ocha.ac.jp}
\author{Eibun Senaha$^{2,3}$}
\email{eibunsenaha@vlu.edu.vn}
\affiliation{$^1$Graduate school of Humanities and Sciences, Ochanomizu University, Tokyo 112-8610, Japan}
\affiliation{$^2$Subatomic Physics Research Group, Science and Technology Advanced Institute, Van Lang University, Ho Chi Minh City, Vietnam}
\affiliation{$^3$Faculty of Applied Technology, School of Technology, Van Lang University, Ho Chi Minh City, Vietnam}

\date{\today}

\begin{abstract}
We study the possibility of electroweak baryogenesis in the standard model with a complex scalar field, focusing mainly on a degenerate scalar scenario. In our setup, CP violation is provided by dimensional-5 Yukawa interactions involving the complex scalar field. In contrast to previous studies in the literature, we exemplify a case in which a complex phase in the singlet scalar potential is transmitted to the fermion sector via the higher-dimensional operators and drives BAU. We point out that an electric dipole moment of the electron can be suppressed due to the Higgs mass degeneracy and the presence of a new electron Yukawa coupling. Thus, viable parameter space for electroweak baryogenesis is still wide open for the latest experimental bound set by the JILA Collaboration.
\end{abstract}


\maketitle


\section{Introduction}\label{sec:intro}
In the standard model (SM), an explanation of a baryon asymmetry of the Universe (BAU) via electroweak baryogenesis (EWBG) mechanism~\cite{Kuzmin:1985mm,Rubakov:1996vz,*Funakubo:1996dw,*Riotto:1998bt,*Trodden:1998ym,*Bernreuther:2002uj,*Cline:2006ts,*Morrissey:2012db,*Konstandin:2013caa,*Senaha:2020mop} is excluded due to a lack of a strong first-order electroweak phase transition (EWPT)~\cite{Kajantie:1996mn,*Rummukainen:1998as,*Csikor:1998eu,*Aoki:1999fi} and insufficient CP violation~\cite{Gavela:1993ts,*Gavela:1994dt,*Huet:1994jb,*Konstandin:2003dx}. Despite its failure, the mechanism is still attractive from the viewpoint of testability, and the EWBG possibility in various models has been actively investigated in light of experiments, such as the Large Hadron Collider and electric dipole moment (EDM) of the electron ($d_e$). One viable scenario compatible with the current LHC data is the so-called \textit{degenerate scalar scenario} in which new scalar masses are close to 125 GeV. Such a scenario can be realized in the SM with a complex scalar (CxSM) and comprehensively studied in connection with dark matter (DM) physics, where it is shown that a spin-independent DM cross section with nucleons is suppressed thanks to the Higgs mass degeneracy~\cite{Abe:2021nih}. Moreover, the scenario can accommodate the strong first-order EWPT though the suppression mechanism for the DM cross section turns out to be another kind~\cite{Cho:2021itv}.

In the CxSM, even though complex phases can, in principle, exist in the scalar potential, they cannot be the sources for BAU since the SU(2) singlet scalar field does not couple to SM fermions directly. The simplest way to get around this problem is to introduce higher-dimensional Yukawa interactions containing the singlet scalar field. If the coefficients of the operators are complex, pseudoscalar interactions would be induced, driving EWBG~\cite{Espinosa:2011eu,Cline:2012hg,Jiang:2015cwa,Cline:2021iff}. 
If the coefficients happen to be real, on the other hand, the CP violation relevant to EWBG should result from the complex phase of the scalar potential. Ref.~\cite{Cho:2022our} shows that the strength of the first-order EWPT could be weakened by the complex phase of the scalar potential, but one could still have the strong first-order EWPT compatible with EWBG. On the other hand, the BAU estimate is not conducted there and is left for future work.

Experimental searches for CP violation beyond the SM are essential for probing the EWBG possibility.
Currently, the electron EDM is the most sensitive to CP violation. In 2018, ACME Collaboration placed an upper bound on $d_e$ as $|d_e^{\text{ACME}}| < 1.1 \times 10^{-29}~e~\text{cm}$ at 90\%~\textrm{C.L.}~\cite{ACME:2018yjb} and in 2022, JILA Collaboration further improved the bound as $|d_e^{\text{JILA}}| < 4.1 \times 10^{-30}~e~\text{cm}$ at 90\%~\textrm{C.L.}~\cite{Roussy:2022cmp}. While maintaining the BAU, some suppression mechanisms should be present to avoid such an unprecedentedly tight EDM bound (for cancellation mechanisms, see, e.g., Refs.~\cite{Fuyuto:2019svr,Kanemura:2020ibp}).

In this letter, we investigate the EWBG feasibility in the CxSM with higher-dimensional operators. In particular, we consider cases where the complex phase exists in the scalar potential, which is transmitted to the SM fermion sector via the dimension-5 Yukawa interactions with and without complex coefficients.
Our study shows that the complex phase of the scalar potential yields the right ballpark value for BAU without resorting to the complex coefficients of the higher-dimensional operators. On the other hand, $d_e$ can be suppressed by the presence of the Higgs mass degeneracy and new electron Yukawa coupling, thus evading the latest upper bound from the JILA experiment.

\section{Model}\label{sec:model}
The CxSM is the extension of the SM by adding a complex SU(2) singlet scalar field ($S$)~\cite{Barger:2008jx,*Barger:2010yn,*Gonderinger:2012rd}. 
In the most general scalar potential, there are 5 real parameters and 8 complex parameters.
As a first step toward the general analysis, we take a principle of minimality to simplify our analysis, which is also employed in our previous work~\cite{Cho:2021itv,Cho:2022our}.\footnote{By the minimality, we mean that the number of the parameters in the scalar potential after imposing the global U(1) symmetry is the smallest under the conditions of no massless Nambu-Goldstone boson, domain wall problems, and no breaking of renormalization.}
The scalar potential we consider in this work is given by
\begin{align}
V_0(H, S) &= \frac{m^2}{2}H^\dagger H+\frac{\lambda}{4}(H^\dagger H)^2
	+\frac{\delta_2}{2}H^\dagger H|S|^2+\frac{b_2}{2}|S|^2 \nonumber\\
&\quad+\frac{d_2}{4}|S|^4
	+\bigg(a_1S+\frac{b_1}{4}S^2+{\rm H.c.}\bigg),
\end{align}
where
\begin{align}
H(x) &=
	\left(
		\begin{array}{c}
		G^+(x) \\
		\frac{1}{\sqrt{2}}\big(v+h(x)+iG^0(x)\big)
		\end{array}
	\right),\\
S(x) &= \frac{1}{\sqrt{2}}\left(v_S^r+iv_S^i  + s(x)+i\chi(x) \right).
\end{align}
Without the $a_1$ and $b_1$ terms, $V_0$ has a global U(1) symmetry, and a massless Nambu-Goldstone boson would appear if the symmetry is spontaneously broken. Moreover, $a_1$ is necessary to break a $Z_2$ symmetry $S\to -S$, which dodges a domain wall problem. If the scalar sector preserves CP, $V_0$ is invariant under the transformation $\chi\to-\chi$, and $\chi$ could be DM. However, as investigated in Ref.~\cite{Cho:2021itv}, the DM relic abundance is too small in the parameter space where EWPT is strong first order. We, therefore, degrade $\chi$ to an ordinary unstable particle by allowing CP violation as needed for EWBG.
While both $a_1$ and $b_1$ can be complex, their relative phase is only physical. As our convention, only $a_1$ is treated as the complex parameter and parametrized as $a_1=a_1^r+ia_1^i$. 

In this setup, the tadpole (minimization) conditions for $h$, $s$, and $\chi$ are respectively given by
\begin{align}
\left\langle \frac{\partial V_0}{\partial h}\right\rangle & = 
v\left[
\frac{m^2}{2} + \frac{\lambda}{4}v^2+\frac{\delta_2}{4}|v_S|^2 
\right]= 0,\label{tad_h}\\
\left\langle \frac{\partial V_0}{\partial s}\right\rangle & = 
v_S^r
\left[
\frac{b_2+b_1}{2}+\frac{\delta_2}{4}v^2+\frac{d_2}{4}|v_S|^2+\frac{\sqrt{2}a_1^r}{v_S^r}
\right]
 = 0,\label{tad_s} \\
\left\langle \frac{\partial V_0}{\partial \chi}\right\rangle & = 
v_S^i
\left[
\frac{b_2-b_1}{2}+\frac{\delta_2}{4}v^2+\frac{d_2}{4}|v_S|^2-\frac{\sqrt{2}a_1^i}{v_S^i}
\right]
= 0,\label{tad_chi}
\end{align}
where the symbol $\langle\cdots \rangle$ denotes that the fluctuation fields are taken zero after their derivatives, and $|v_S|^2=v_S^{r2}+v_S^{i2}$. 
After imposing the tadpole conditions, the mass matrix in the basis $(h, s, \chi)$ is cast into the form
\begin{align}
\mathcal{M}_S^2= 
\begin{pmatrix}
\frac{\lambda}{2}v^2 & \frac{\delta_2}{2}vv_S^r & \frac{\delta_2}{2}vv_S^i \\
\frac{\delta_2}{2}vv_S^r  & \frac{d_2}{2}v_S^{r2}-\frac{\sqrt{2}a_1^r}{v_S^r} & \frac{d_2}{2}v_S^rv_S^i \\
 \frac{\delta_2}{2}vv_S^i & \frac{d_2}{2}v_S^rv_S^i & \frac{d_2}{2}v_S^{i^2}+\frac{\sqrt{2}a_1^i}{v_S^i}
\end{pmatrix}, \label{M}
\end{align}
which is diagonalized by an orthogonal matrix $O$ as $O^T\mathcal{M}_S^2O = \text{diag}(m_{h_1}^2, m_{h_2}^2, m_{h_3}^2)$. We parametrize the matrix $O$ as
\begin{align}
O & = 
\begin{pmatrix}
1 & 0 & 0 \\
0 & c_3 & -s_3 \\
0 & s_3 & c_3
\end{pmatrix}
\begin{pmatrix}
c_2 & 0 & -s_2  \\
0 & 1 & 0 \\
s_2 & 0 & c_2
\end{pmatrix}
\begin{pmatrix}
c_1 & -s_1 & 0 \\
s_1 & c_1 & 0 \\
0 & 0 & 1
\end{pmatrix},
\end{align}
where $s_i=\sin\alpha_i$ and $c_i=\cos\alpha_i~(i=1,2,3)$.
In our work, the 8 original parameters $\{m^2$, $\lambda$, $\delta_2$, $b_2$, $d_2$, $a_1^r$, $a_1^i$, $b_1\}$ are converted to $\{v$, $v_S^r$, $v_S^i$, $m_{h_1}$, $m_{h_2}$, $m_{h_3}$, $\alpha_1$, $\alpha_2 \}$ using the tadpole conditions together with the mass condition.
We note that $a_1^i$ is given by
\begin{align}
a_1^i & = \frac{v_S^i}{\sqrt{2}}
\sum_iO_{3i}\left(O_{3i}-O_{2i}\frac{v_S^i}{v_S^r}\right)m_{h_i}^2,
\end{align}
which implies that $a_1^i=0$ if $v_S^i=0$, i.e., the explicit CP violation must be associated with the spontaneous CP violation, but not vice versa. 

The Higgs couplings to fermions ($f$) and gauge bosons ($V=Z, W^\pm$) are defined as
\begin{align}
\mathcal{L}_{h_i\bar{f}f} &= -\frac{m_f}{v}\sum_{i=1}^3\kappa_{if} h_i\bar{f}f, \\
\mathcal{L}_{h_iVV} &= \frac{1}{v}\sum_{i=1}^3\kappa_{iV} h_i(m_Z^2Z_\mu Z^\mu+2m_W^2W_\mu^+W^{-\mu}),
\end{align}
where $\kappa_{if}=O_{1i}$ and $\kappa_{iV}=O_{1i}$. Note that the presence of the complex parameters in the scalar potential does not give rise to pseudoscalar coupling in $\mathcal{L}_{h_i\bar{f} f}$, meaning that EWBG is not driven in this setup. To circumvent this issue, we introduce dimensional-5 operators. The relevant terms in the following discussion are
\begin{align}
-\mathcal{L}_{h_i\bar{f}f}^{\text{dim.5}} &\ni \bar{q}_{L}\tilde{H}\left(y_t+\frac{c_t}{\Lambda}S\right)t_{R}+\bar{\ell}_{L}H\left(y_e+\frac{c_e}{\Lambda}S\right)e_{R} \nonumber \\
&\quad +\text{H.c.}, \label{EFT-dim5}
\end{align}
where $q_L$ denotes the up-type left-handed quark doublet of the third generation, while $\ell_L$ is the down-type left-handed lepton doublet of the first generation. $t_R$ and $e_R$ are the right-handed top and electron, respectively. 
$\Lambda$ is a cutoff scale and $\tilde{H}=i\tau^2H^*$ with $\tau^2$ representing the second Pauli matrix.
$y_t$ and $y_e$ are the top and electron Yukawa couplings in the SM, respectively, while $c_t$ and $c_e$ are general complex parameters. For later use, we parametrize $c_f = |c_f|e^{i\phi_f}=c_f^r+ic_f^i,~f=t,e$.
As shown below, $c_e$ could be pivotal in suppressing the electron EDM. 

Let us redefine the Higgs couplings to the fermions in the presence of the dimension-5 operators as
\begin{align}
\mathcal{L}_{h_i\bar{f}f}&= -\sum_{i=1}^3 h_i\bar{f}\Big(g_{h_i\bar{f}f}^S+ig_{h_i\bar{f}f}^P\gamma_5\Big)f,
\end{align}
where
\begin{align}
g_{h_i\bar{f}f}^S & = \frac{1}{\sqrt{2}}
\left[
	y_fO_{1i}+\frac{v}{\sqrt{2}\Lambda}(c_f^rO_{2i}-c_f^iO_{3i})
\right], \\
g_{h_i\bar{f}f}^P & = \frac{v}{\sqrt{2}\Lambda}(c_f^rO_{3i}+c_f^iO_{2i}).
\end{align}
As seen, the pseudocouplings $g_{h_i\bar{f}f}^P$ exist because of the dimension-5 operators, and $\chi$ is now interpreted as the pseudoscalar.  

Our primary interest is the case in which the complex phase in the scalar potential is the only source for the CP violation that drives EWBG. Secondarily, to what extent complex $c_t$ and $c_e$ can change the former result.  
In what follows, we consider the 2 cases:
\begin{itemize}
\item[1.] Both $c_t$ and $c_e$ are real
\item[2.] Both $c_t$ and $c_e$ are complex
\end{itemize}
We make a comment on a case in which $c_t$ is complex while $c_e$ is real at the end of Sec.~\ref{sec:results}.

Before closing this section, we briefly describe the degenerate scalar scenario that can mimic the SM. 
For illustration, we consider a process $gg\to h_i \to VV^*$. Since $|m_{h_i}-m_{h_j}|>(m_{h_i}\Gamma_{h_i}+m_{h_j}\Gamma_{h_j})/(m_{h_i}+m_{h_j})$ in our benchmark points, where $\Gamma_{h_i}$ are the total decay width of $h_i$, we can use a narrow decay width approximation~\cite{Fuchs:2014ola,Das:2017tob}. With the approximation, the cross section normalized by the SM value is cast into the form
\begin{align}
\frac{\sigma_{gg\to h_i \to VV^*}}{\sigma_{gg\to h_i \to VV^*}^{\text{SM}}}
\simeq 1+ \frac{v^2|c_t|^2}{\Lambda^2y_t^2},
\end{align}
where we have used $\Gamma_{h_i}\simeq \kappa_{iV}^2\Gamma_{h}^{\text{SM}}$ with $\Gamma_{h}^{\text{SM}}$ being the total decay width of the SM Higgs boson.
For $|c_t|=y_t$ and $\Lambda=1.0$ TeV, the deviation from the SM value would be about $6\%$,
which is still consistent with the current LHC data~\cite{ATLAS:2022vkf,CMS:2022dwd}.\footnote{Note that deviations of other processes such as the Higgs decay to diphoton are also $\mathcal{O}(\frac{v^2|c_t|^2}{\Lambda^2 y_t^2})\sim 6\%$ in our study, which is consistent with the current LHC data~\cite{ATLAS:2022vkf,CMS:2022dwd}.}
While somewhat lower $\Lambda$ could be allowed experimentally, detailed collider analysis would be required for that, 
and we do not pursue this possibility in the current work. We have confirmed that our conclusion does not change even when $\Lambda=0.5$ TeV.

Currently, experimental constraints on the Higgs total decay width are $\Gamma_h^{\text{exp}}<14.4$ MeV (ATLAS~\cite{ATLAS:2018jym}) and $\Gamma_h^{\text{exp}}=3.2^{+2.4}_{-1.7}$ MeV (CMS~\cite{CMS:2022ley}), which are not precise enough to provide a valuable constraint to our scenario.

\section{Electroweak baryogenesis}\label{sec:ewbg}
We are following closely the work of Refs.~\cite{Cline:2000nw,Fromme:2006wx,Cline:2020jre}, derive the semiclassical force in the presence of the CP violation discussed in the previous section. 
The Yukawa interaction with a spacetime-dependent complex mass is defined as
\begin{align}
\mathcal{L}_Y=\bar{f}\big(i\dsla-m_f(x)P_R-m_f^*(x)P_L\big)f,
\end{align}
where $\dsla=\gamma^\mu\partial_\mu$.
Since the thickness of the bubble wall is much smaller than that of the radius, we can approximate it as
a planner. In this case, the spacetime dependence of $m_f$ is only $z$ which is the coordinate of the perpendicular to the wall.

From the above Yukawa Lagrangian, the equation of motion is given by
\begin{align}
\big(i\partial_z\kern-2ex /~-m_f(z)P_R-m_f^*(z)P_L\big)f=0,
\end{align}
where $m_f(z)\equiv |m_f(z)|e^{i\theta_f(z)}$. The semiclassical force is found to be
\begin{align}
F_z&=-\frac{(|m_f|^2)'}{2E}\pm s\frac{(|m_f|^2\theta_f')'}{2E_0E_{0z}},
\end{align}
where
\begin{align}
E & = E_0\mp s\frac{|m_f|^2\theta'_f}{2E_0E_{0z}}, \\
E_0&=\sqrt{p^2_x+p^2_y+p^2_z+|m_f|^2}, \\
E_{0z}&=\sqrt{p^2_z+|m_f|^2}.
\end{align}
The upper and lower signs correspond to particles and antiparticles, respectively.
We also note that particles with opposite spin receive the opposite CP-violating force.
The nonzero momenta parallel to the wall can enhance the CP-violating part, as referred to by Ref.~\cite{Fromme:2006wx}.

In our case, the top mass during EWPT has the form
\begin{align}
m_t(z) = \frac{\rho(z)}{\sqrt{2}}\left(y_t+\frac{c_t}{\sqrt{2}\Lambda}\big(\rho_S^r(z)+i\rho_S^i(z)\big)\right),
\end{align}
where $\rho(z)$, $\rho_S^{r}(z)$, and $\rho_S^{i}(z)$ are the bubble wall profiles parametrized as $\langle H\rangle^T=(0~\rho(z)/\sqrt{2})$, $\langle S\rangle = (\rho_S^r(z)+i\rho_S^i(z))/\sqrt{2}$, while the phase $\theta_t(z)$ is expressed as
\begin{align}
\theta_t(z) & = \tan^{-1}\left(\frac{ c_t^r\rho_S^i(z)+c_t^i\rho_S^r(z) }{\sqrt{2}\Lambda+c_t^r\rho_S^r(z)-c_t^i\rho_S^i(z) }\right).
\label{theta_t}
\end{align}
The detail of the bubble wall calculations is given in Ref.~\cite{Cho:2022our}.

After solving transport equations, one can find the baryon-to-photon ratio ($\eta_B$) as~\cite{Cline:2020jre}
\begin{align}
\eta_B = \frac{405\Gamma_{\text{sph}}^{\text{sym}}}{4\pi^2 \gamma_wv_wg_*(T)T}
\int_0^\infty dz~\mu_{B_L}\exp\left(-\frac{45\Gamma_{\text{sph}}^{\text{sym}}z}{4\gamma_wv_w}\right),
\label{etaB}
\end{align}
where $\mu_{B_L}$ denotes a chemical potential for the left-handed baryon number, $g_*(T)(=108.75)$ is the degrees of freedom of the relativistic particles in the thermal bath, $\Gamma_{\text{sph}}^{\text{sym}}(=1.0\times 10^{-6}T$~\cite{Cline:2020jre}) is the sphaleron rate in the symmetric phase, $v_w(=0.1)$ is the wall velocity, and $\gamma_w=1/\sqrt{1-v_w^2}$. We set $T$ to a nucleation temperature $T_N=66.847$ GeV for $\eta_B$.
Using Eq.~(\ref{etaB}), we estimate $\eta_B$ and compare with the observed values, $\eta_B^{\text{BBN}}=(5.8-6.5)\times 10^{-10}$ at 95\% CL from bigbang nucleosynthesis and $\eta_B^{\text{CMB}}=(6.105\pm 0.055)\times 10^{-10}$ at 95\% CL  from comic microwave background~\cite{ParticleDataGroup:2022pth}. 

\section{Electric dipole moments}\label{sec:edms}
\begin{figure}[t]
\begin{center}
\includegraphics[width=4cm]{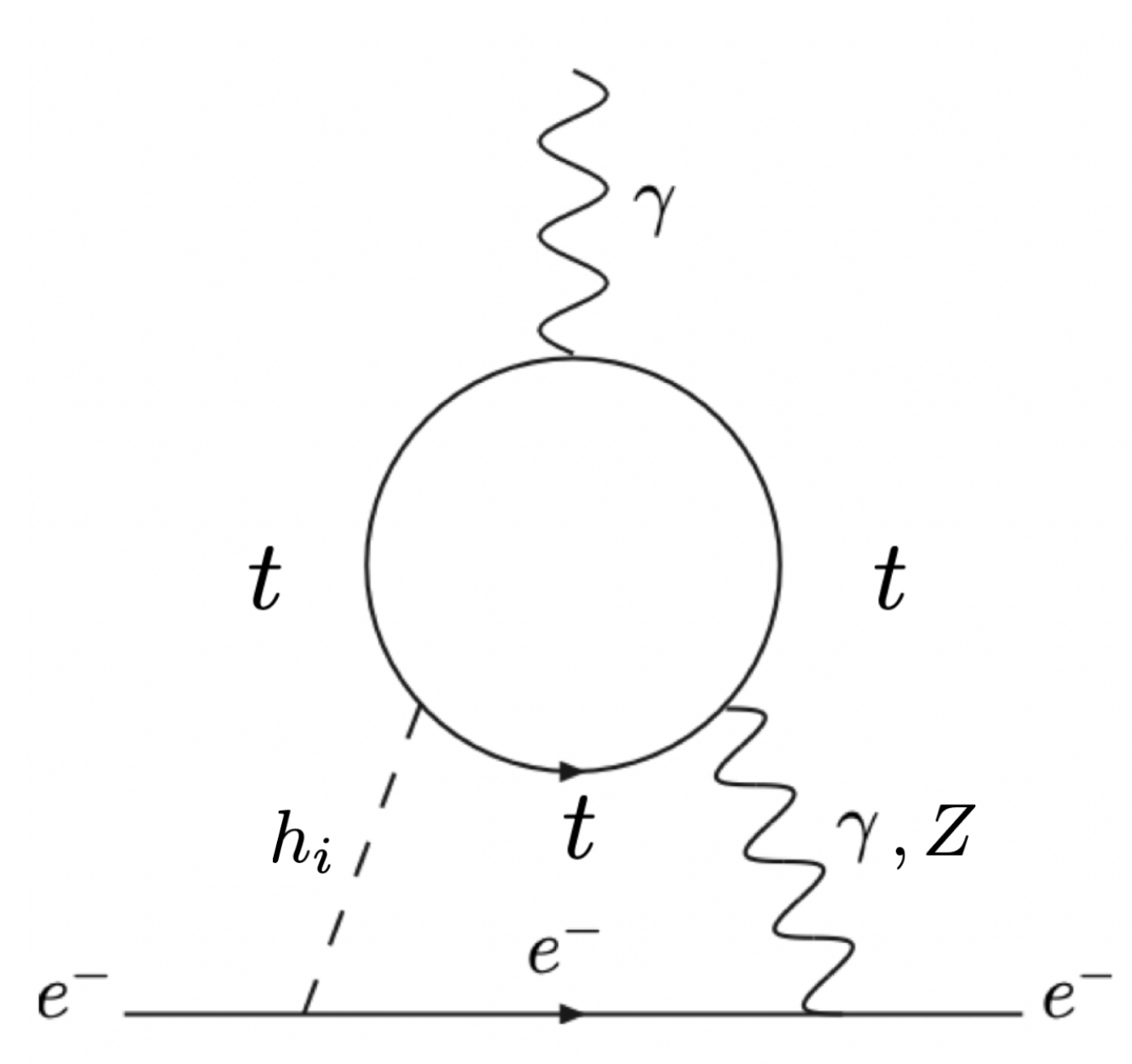}
\includegraphics[width=4cm]{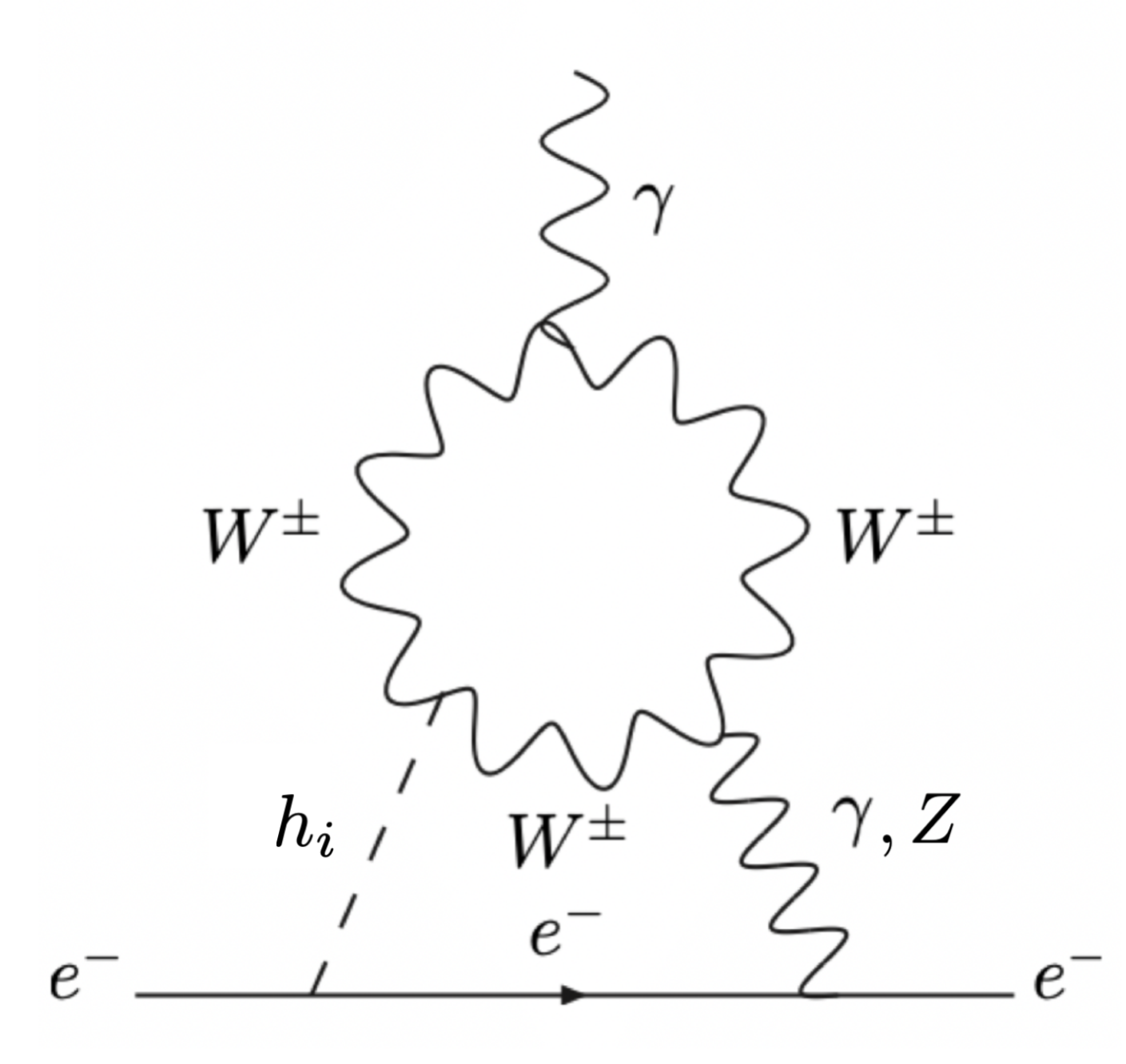}
\caption{Dominant two-loop contributions to $d_e$. The left diagrams are denoted as $(d_e^{h\gamma})_t$ and $(d_e^{hZ})_t$, while the right ones as  $(d_e^{h\gamma})_W$ and $(d_e^{hZ})_W$. 
}
\label{fig:BZ}
\end{center}
\end{figure}

EDMs, especially the electron EDM, severely constrain the magnitude of CP violation.
The latest upper bounds on $|d_e|$ from the ACME and JILA experiments are, respectively, given by~\cite{ACME:2018yjb,Roussy:2022cmp}
\begin{align}
|d_e^{\textrm{ACME}}| &< 1.1 \times 10^{-29}~e~\text{cm}~(90\%~\textrm{C.L.})\label{de_ACME18}, \\
|d_e^{\textrm{JILA}}| &< 4.1 \times 10^{-30}~e~\text{cm}~(90\%~\textrm{C.L.}).\label{de_JILA22}
\end{align}
In our model, dominant corrections to $d_e$ come from the so-called Barr-Zee diagrams~\cite{Barr:1990vd}. We decompose them into two parts
\begin{align}
d_e = d_e^{t}+d_e^{W},
\end{align}
where $d_e^{t}=(d_e^{h\gamma})_t+(d_e^{hZ})_t$ and $d_e^{W}=(d_e^{h\gamma})_W+(d_e^{hZ})_W$ with the subscripts of the parentheses representing the particle running in the upper loop in the Barr-Zee diagrams, as depicted in Fig.~\ref{fig:BZ}.

The top-loop contributions to $d_e$ in the degenerate mass limit becomes
\begin{align}
\frac{(d_e^{h\gamma})_t}{e}
& \simeq  \frac{\alpha_{\text{em}}|c_{t}||c_{e}|\sin(\phi_{e}-\phi_{t})v^2}{24\pi^3m_t\Lambda^2}
\Big[ f(\tau_{th})-g(\tau_{th}) \Big], 
\label{de_hgam_t}
\end{align}
where $\tau_{th} = m_t^2/m_h^2$ with $m_h\equiv m_{h_1}=m_{h_2}=m_{h_3}$, and $c_{t,e}=|c_{t,e}|e^{i\phi_{t,e}}$.
$f(\tau_{th})$ and $g(\tau_{th})$, are the loop functions defined in Refs.~\cite{Barr:1990vd}. In our convention, $e$ represents the positron charge. Eq.~(\ref{de_hgam_t}) implies that $(d_e^{h\gamma})_t$ vanishes when $\phi_t=\phi_e+n\pi$ with $n$ being the integer, let alone $c_t$ and $c_e$ are both real.

The $W$-loop contributions to $d_e$ are induced by the complex $c_e$, which have the form 
\begin{align}
\frac{(d_{e}^{h\gamma})_W}{e}
&=-\sum_{i=1}^3\frac{\alpha_{\text{em}}^2vC^{h_i\gamma}_W}{32\pi^2s_W^2m_W^2}\mathcal{J}^\gamma_W(m_{h_i}),
\end{align}
where $\mathcal{J}_W^\gamma(m_{h_i})$ denotes the loop function~\cite{Abe:2013qla}, and one can find
\begin{align}
\lefteqn{\sum_{i=1}^3C^{h_i\gamma}_W\mathcal{J}^\gamma_W(m_{h_i}) } \nonumber\\
 &= \frac{vc_e}{2\Lambda} 
 \big(O_{12}O_{32}\Delta^{(21)}\mathcal{J}^\gamma_W+O_{13}O_{33}\Delta^{(31)}\mathcal{J}^\gamma_W \big),
\end{align}
where $\Delta^{(ij)}\mathcal{J}^\gamma_W=\mathcal{J}^\gamma_W(m_{h_i})-\mathcal{J}^\gamma_W(m_{h_j})$.
Therefore, regardless of $c_e$, $(d_{e}^{h\gamma})_W$ vanishes when $m_{h_1}=m_{h_2}=m_{h_3}$.

Similarly, we can obtain the same vanishing conditions for $(d_e^{hZ})_t$ and $(d_e^{hZ})_W$.
The conditions for the vanishing $d_e^t$ and $d_e^W$ are summarized in Table~\ref{tab:suppression}.

\begin{table}[t]
\centering
  \begin{tabular}{|c|c|c|}  \hline
    & $d_e^t$ & $d_e^W$ \\ \hline \hline
    Real $c_t$ and $c_e$ & $m_{h_i}=m_{h_j}$ & $m_{h_i}=m_{h_j}$  \\ \hline
    Complex $c_t$ and $c_e$ & $m_{h_i}=m_{h_j}$ and $\phi_t=\phi_e\pm n\pi$ & $m_{h_i}=m_{h_j}$  \\ \hline
    \end{tabular}
    \caption{Conditions for the vanishing electron EDM.}
  \label{tab:suppression}
\end{table}

\begin{figure}[t]
\begin{center}
\includegraphics[width=7.3cm]{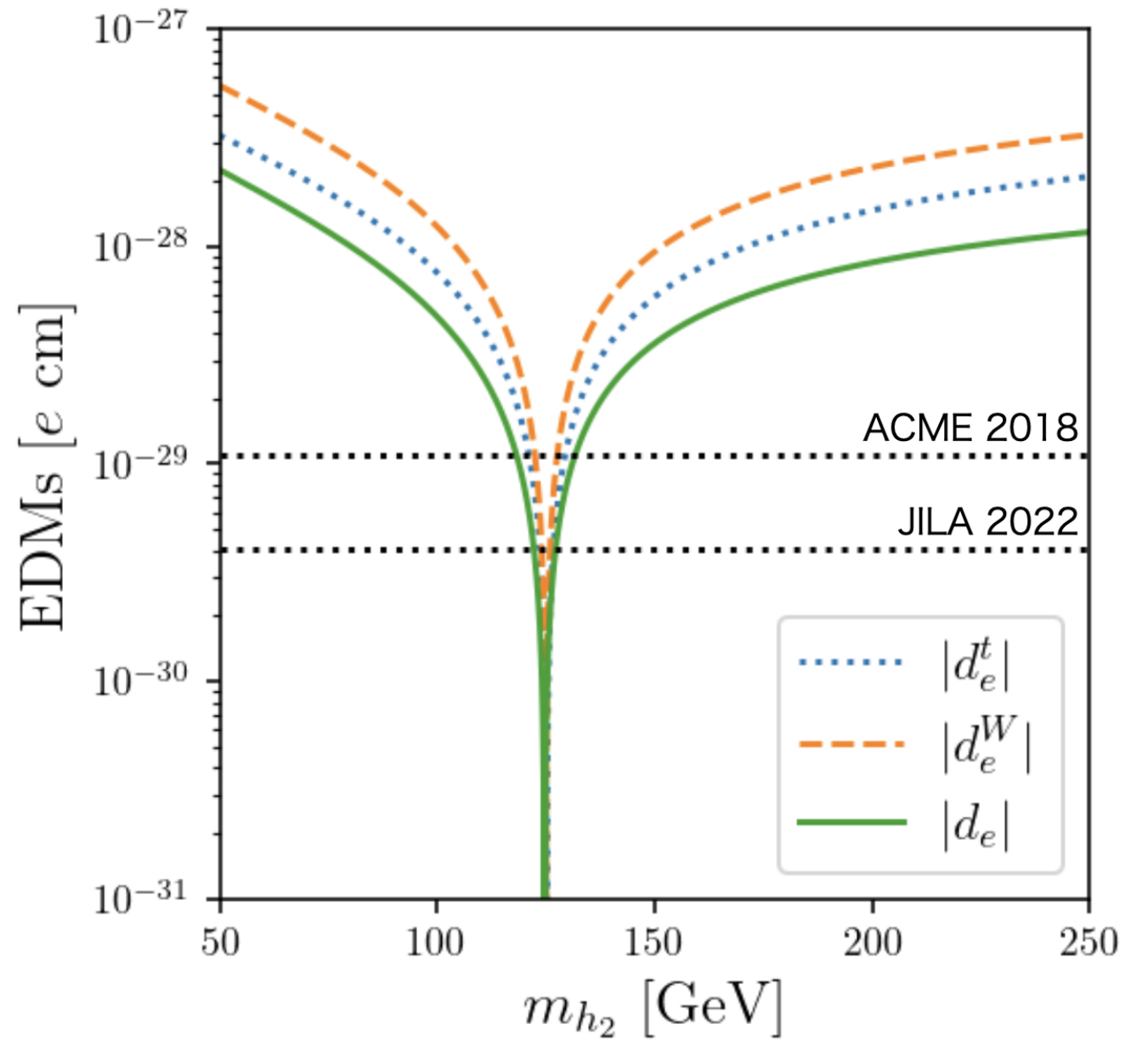}
\caption{The electron EDM as a function of $m_{h_2}$ in the case that $|c_t|=y_t$, $|c_e|=y_e$, $\phi_t=\phi_e=0$, and $\Lambda=1.0$ TeV. 
We take the parameter set given in Table~\ref{tab:BP1} while $m_{h_2}$ is treated as the free parameter.
 Here, $d_e^t$ and $d_e^W$ are the two-loop contributions to the electron EDM, depicted as the left and right diagrams in Fig.~\ref{fig:BZ}, respectively.
}
\label{fig:EDM_real}
\end{center}
\end{figure}

To see the suppression behavior numerically, a typical example is given here.
The input parameters are summarized in Table~\ref{tab:BP1} but with $m_{h_2}$ being free.
In this example, we take $\Lambda=1.0$ TeV, $|c_t|=y_t$, $|c_e|=y_e$, $\phi_t=\phi_e=0$, and thus CP violation arises from the nonzero $v_S^i$.
Fig.~\ref{fig:EDM_real} shows $|d_e|$ (green solid line) and its details $|d_e^t|$ (blue dotted line) and $|d_e^W|$ (orange dashed line) against $m_{h_2}$. The upper dotted horizontal line denotes the experimental bound of ACME, while the lower one represents the JILA bound. As discussed above, the both $|d_e^t|$ and $|d_e^W|$ would be suppressed as $m_{h_2}$ approaches $125~\text{GeV}(=m_{h_1})$, evading ACME and JILA constraints. This example clearly illustrates that the degenerate scalar scenario simultaneously provides an exquisite parameter space compatible with the LHC and the electron EDM data. 

\section{Numerical results and discussions}\label{sec:results}

\begin{table}[t]
\centering
  \begin{tabular}{|c|c|c||c|c|}  \hline
    & $\eta_B/10^{-10}$ & $|d_e|/10^{-30}$ & $d_e^t/10^{-30}$ & $d_e^W/10^{-30}$ \\ \hline
    $\Lambda=1.0$ [TeV] & $1.16$ & $1.15$ & $3.14$ & $-4.29$  \\ \hline
    $\Lambda=1.5$ [TeV] & $0.797$ & $0.77$  & $2.09$ & $-2.86$ \\ \hline
    $\Lambda=2.0$ [TeV] & $0.606$ & $0.57$ & $1.57$ & $-2.15$ \\ \hline
    \end{tabular}
    \caption{Summary of $\eta_B$ and $|d_e|$ in the case of $|c_t|=y_t$, $|c_e|=y_e$, and $\phi_t=\phi_e=0$. The electron EDM is given in units of $e$ cm.}
  \label{tab:bau-edm}
\end{table}

\begin{table*}[t]
\center
\begin{tabular}{|c|c|c|c|c|c|c|c|c|}
\hline
Inputs & $v$ [GeV] & $v_S^r$ [GeV] & $v_S^i$ [GeV] & $m_{h_1}$ [GeV] & $m_{h_2}$ [GeV] & $m_{h_3}$ [GeV] & $\alpha_1$ [rad] & $\alpha_2$ [rad] \\ \hline
 & 246.22 & 0.6 & $-0.3$ & 125.0 & 124.0 & 124.5 & $\pi/4$ & 0.0  \\ \hline
Outputs & $m^2$ & $b_2$ [GeV$^2$] & $b_1$ [GeV$^2$] & $\lambda$ & $\delta_2$ & $d_2$ & $a_1^r$ [GeV$^3$] & $a_1^i$ [GeV$^3$]  \\ \hline
 & $-(124.5)^2$ & $-(121.2)^2$ & $-7.717\times 10^{-12}$& 0.511 &1.51 & 1.111 & $-(18.735)^3$ & $-(14.870)^3$ \\ 
\hline
\end{tabular}
\caption{Inputs and outputs in our benchmark. In this case, $\alpha_3=0.464$ radians, and the Higgs coupling modifiers are $\kappa_{1}=0.711$, $\kappa_2=-0.711$, and $\kappa_3=0.0$.}
\label{tab:BP1}
\end{table*}

As studied in Ref.~\cite{Cho:2022our}, $0.3\lesssim v_S^i \lesssim 0.5$ is the range where the first-order EWPT is strong enough to suppress baryon-changing processes and bubble nucleation happens. 
Since the first-order EWPT is driven by a tree-level potential barrier, its strength would remain unchanged even after including the dimension-5 Yukawa operators (\ref{EFT-dim5}). We take a parameter set BP1 adopted in Ref.~\cite{Cho:2022our} for illustrative purposes but with the sign of $v_S^i$ being flipped. The inputs and outputs are summarized in Table~\ref{tab:BP1}. 
Regarding $c_f~(f=t,e)$, we set $|c_f| = y_f$ and take $\phi_f$ as the free parameters.

In the case of $\phi_t=\phi_e=0$, CP violation solely comes from the scalar potential. With this CP violation, we calculate the BAU in the cases of $\Lambda=1.0$, 1.5, and 2.0 TeV, respectively. The results are summarized in Table~\ref{tab:bau-edm}, where $|d_e|$ and its details are also shown. One can see that the $\Lambda=1.0$ TeV case yields $\eta_B=\mathcal{O}(10^{-10})$, while the other two cases provide the smaller $\eta_B$ to some extent. Even though the obtained values of $\eta_B$ are somewhat insufficient for explaining the observed one, we make no strong claims about the numbers since the perturbative calculations of EWPT and BAU employed in this work are generally subject to significant theoretical uncertainties. Further theoretical improvements should be left to future work.

\begin{figure}[t]
\begin{center}
\includegraphics[width=7cm]{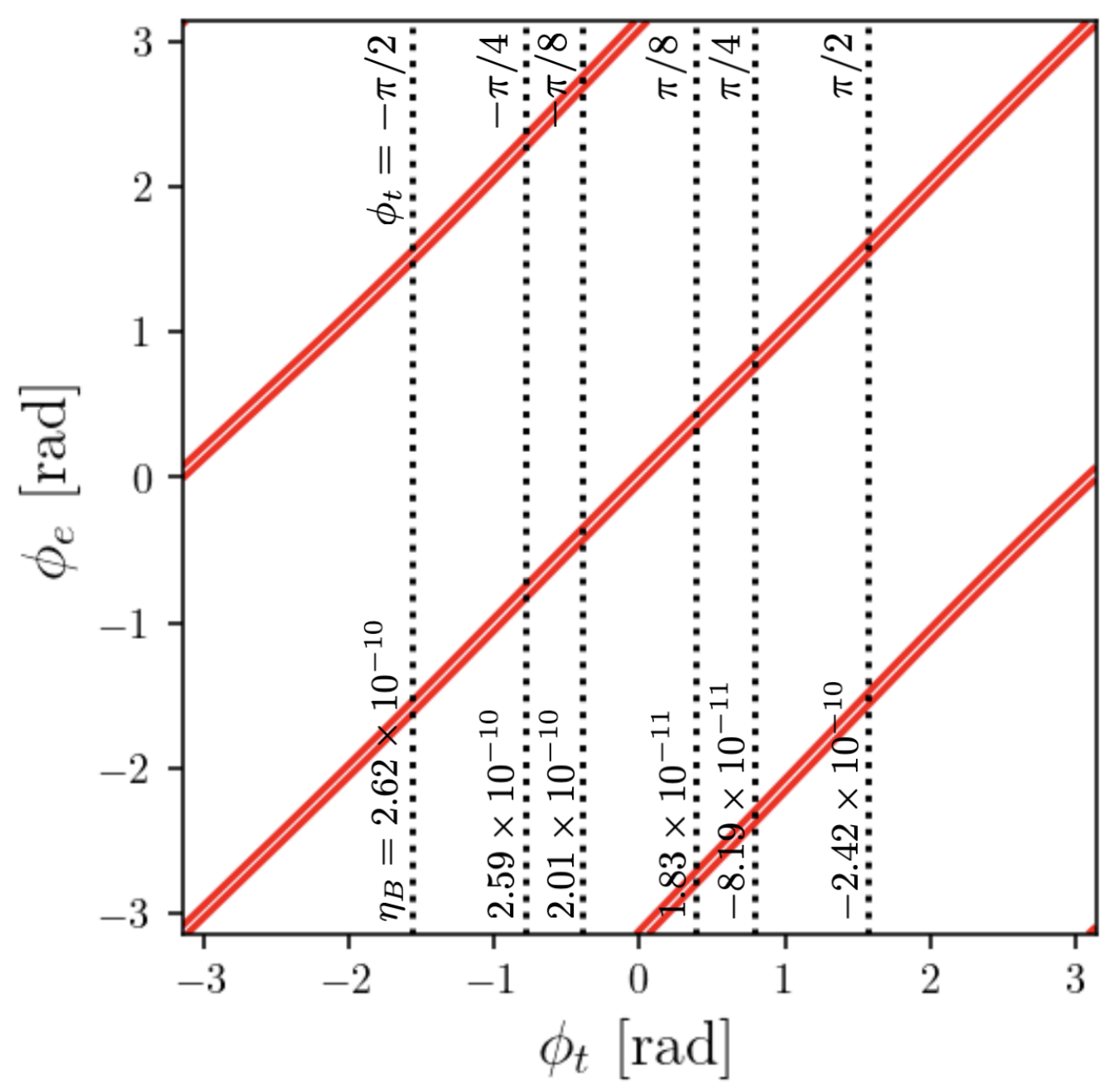}
\caption{$\eta_B$ and $|d_e|$ are shown, where $\Lambda=1.0$ TeV, $|c_t|=y_t$, and $|c_e|=y_e$ are taken. The narrow bands by the diagonal lines satisfy $|d_e|<|d_e^{\text{JILA}}|$.
}
\label{fig:EDM_complex}
\end{center}
\end{figure}

Now, we discuss the case of complex $c_t$ and $c_e$.
In this case, there are three sources for CP violation, and $v_S^i$ and $c_t$ are responsible for EWBG. 
Fig.~\ref{fig:EDM_complex} displays $\eta_B$ and $|d_e|$ in the $(\phi_t, \phi_e)$ plane. 
The vertical dotted lines denotes $\eta_B=2.62\times 10^{-10}$, $2.59\times 10^{-10}$, $2.01\times 10^{-10}$, $1.83\times 10^{-11}$, $-8.19\times 10^{-11}$, and $-2.42\times 10^{-10}$ for $\phi_t=-\pi/2$, $-\pi/4$, $-\pi/8$, $\pi/8$, $\pi/4$, and $\pi/2$, from left to right, respectively. In this benchmark point, $\phi_t=-\pi/2$ gives the largest BAU with the correct sign. 
Compared to the real $\phi_t$ case, $\eta_B$ could get enhanced but not drastically. Slightly short of the correct BAU value may be explained by theoretical uncertainties not considered here. We also show the regions $|d_e|<|d_e^{\text{JILA}}|$ by the diagonal narrow bands in which $\phi_t=\phi_e\pm\pi$ is satisfied. This demonstration clarifies that the successful EWBG parameter space is still wide open in light of the JILA data.

Finally, some comments are noted.
\begin{itemize}
\item One may ask whether the cancellation of the electron EDM can occur in concert with the complex $c_t$ without resorting to the phase alignment with $c_e$. In principle, this can happen. However, this type of cancellation becomes effective only when the scalar masses are not close to each other.
\item Other EDMs such as neutron and Mercury could be significant in exploring this scenario. In doing so, however,  it is necessary to introduce additional new Yukawa couplings of the first-generation quarks. This topic should be studied separately from the present analysis.  
\item Instead of the dimension-5 operators, we could consider dimension-6 Yukawa interactions, such as
\begin{align}
-\mathcal{L}_{h_i\bar{f}f}^{\text{dim.6}} &\ni \bar{q}_{L}\tilde{H}\left(y_t+\frac{c_t'}{\Lambda^2}S^2\right)t_{R} \nonumber\\
&\quad+ \bar{\ell}_{L}H\left(y_e+\frac{c_e'}{\Lambda^2}S^2\right)e_{R}+\text{H.c.}.
\label{EFT-dim6}
\end{align}
From the dimensional analysis, CP violation in this case would be more suppressed than in the dimension-5 operator case. It is found that $\eta_B<1.0\times 10^{-10}$ and $|d_e|<1.0\times 10^{-30}~e~\text{cm}$ for the same parameter set as in the dimension-5 operator case. In this case, the EDM suppressions due to the additional factor $1/\Lambda$ and scalar mass degeneracy are strong enough to avoid the EDM bounds, and the phase alignment $\phi_t=\phi_e+n \pi$ is not necessarily required.
\item In the general scalar potential, we have more complex parameters coming from $S^3$, $SH^\dagger H$, etc.
In such an enlarged parameter space, the EDM cancellation would be more effective, while the BAU may be more enhanced. 
\item Double Higgs production processes are one of the interesting collider signatures of EWBG. 
As mentioned in Sec.~\ref{sec:model}, the modification by the top Yukawa couplings is typically $6\%$. 
On the other hand, the triple Higgs couplings in this model could get large compared to the SM value.
Among all the triple Higgs couplings $\lambda_{h_ih_jh_k}~ (i=1,2,3)$, we find that $\lambda_{h_1h_1h_1}$ is the largest in our benchmark point, which is about 1.4 times larger than that in the SM. Even though the current LHC cannot measure the triple Higgs coupling~\cite{CMS:2022dwd,ATLAS:2022jtk}, future colliders may be capable.
We defer the detailed analysis to future work. 
\end{itemize}

\section{Conclusion}\label{sec:conclusion}

We have studied the possibility of EWBG in the CxSM with the dimension-5 Yukawa interactions. We consider the two cases: one is the case in which CP violation arises only from the scalar potential and propagates to the SM fermion sector by the dimension-5 top Yukawa interaction, and the other is the case where the coefficient of the dimension-5 Yukawa interaction additionally yields CP violation.
It is found that the former leads to $\eta_B=\mathcal{O}(10^{-10})$, and the additional CP violation in the latter helps to increase $\eta_B$ to some extent. Even though the nominal values of $\eta_B$ in our benchmark points are smaller than the observed value by a factor of a few, the deficit might be compensated by theoretical uncertainties that could reside in the perturbative treatments of EWPT and BAU. A more elaborate analysis will be left to future research.

We also investigated the electron EDM in the two cases mentioned above. The electron EDM is suppressed due to the Higgs mass degeneracy, and the ACME and JILA constraints can be evaded for the real $c_t$ and $c_e$ cases. In contrast, in the complex $c_t$ and $c_e$ case, the phase alignment $\phi_t=\phi_e+n\pi$ is additionally needed to be consistent with the experimental bounds.  

In conclusion, the EWBG parameter space in our scenario is still wide open after the recent EDM updates. 

\begin{acknowledgments}
We thank Hiroto Shibuya for the valuable discussions.
The work of C.I. was supported by JST, the establishment of university fellowships towards the creation of science technology innovation, Grant No. JPMJFS2113.
\end{acknowledgments}

\appendix
\section{UV model}
By analogy with the work of Ref.~\cite{Cline:2021iff}, one of the UV models that generate the higher-dimensional operators~(\ref{EFT-dim5}) and (\ref{EFT-dim6}) would be 
\begin{align}
-\mathcal{L} &= \bar{q}_L\tilde{H}Y^uu_R+\bar{q}_L\tilde{H}\lambda_1^uU_R+\bar{U}_L\lambda_2^uu_RS \nonumber\\
&\quad+S\bar{U}_L\lambda_S^uU_R+\bar{U}_LM^uU_R \nonumber \\
&\quad+\bar{\ell}_LHY^ee_R+\bar{\ell}_LH\lambda_1^eE_R+\bar{E}_L\lambda_2^ee_RS \nonumber\\
&\quad  +S\bar{E}_L\lambda_S^eE_R+\bar{E}_LM^eE_R
+\text{H.c.},
\end{align}
where $u_R$ are up-type SM quarks, while $U_{L,R}$ and $E_{L,R}$ are the vector-like (VL) fermions. The omitted down-type quarks can be introduced in the same manner. 
The SM quantum numbers of each field is respectively given by $(\boldsymbol{3},\boldsymbol{1},2/3)$ for $U_{L,R}$, $(\boldsymbol{1},\boldsymbol{1},-1)$ for $E_{L,R}$. In principle, the VL fermions could have multiple flavors, and $\lambda_{1,2,S}^{u,e}$ and $M^{u,e}$ could be complex matrices. As usual, $M^{u,e}$ can be diagonalized by bi-unitary transformations of the VL fermions. However, $\lambda_{1,2,S}^{u,e}$ are general complex matrices. For illustration, we focus on the up-type fermions neglecting off-diagonal flavors and denoting the common mass scale of the VL fermions as $M$. 
By integrating all the VL fermions, one can find 
\begin{align}
-\mathcal{L}_{\text{EFT}} &= \bar{q}_L\tilde{H}
\left[
y_t-\frac{\lambda_1^u\lambda_2^u}{M}S+\frac{\lambda_1^u\lambda_S^u\lambda_2^u}{M^{2}}S^2+\cdots
\right]t_R \nonumber \\
&\quad +\bar{\ell}_LH
\left[
y_e-\frac{\lambda_1^e\lambda_2^e}{M}S+\frac{\lambda_1^e\lambda_S^e\lambda_2^e}{M^{2}}S^2+\cdots
\right]e_R \nonumber \\
&\quad +\text{H.c.}, \label{UVEFTL}
\end{align}
where only the top and electron parts are shown.  Comparing Eq.~(\ref{UVEFTL}) with Eqs.~(\ref{EFT-dim5}) and (\ref{EFT-dim6}), one obtains the following relations:
\begin{align}
\frac{c_f}{\Lambda} = -\frac{\lambda_1^f\lambda_2^f}{M}, \quad
\frac{c_f'}{\Lambda^2} = \frac{\lambda_1^f\lambda_S^f\lambda_2^f}{M^{2}}
\end{align}
for $f=t,e$.

%
\bibliography{refs}
%

\end{document}